\documentstyle[aps,prb,preprint,epsf]{revtex}

\begin{document}


\title {Bose-Einstein condensation of finite number of confined particles}
\author { Wenji Deng$^{a,b}$\ \ \  and \ \ \  P. M. Hui$^{a}$\\
$^{a}$ {\em Department of Physics, The Chinese University of Hong Kong, \\
Shatin, New Territories, Hong Kong.}\\
$^{b}$ {\em Department of Physics, South China University of Technology,\\
Guangzhou 510641, P. R. China.}}

\maketitle

\begin {abstract}
The partition function and specific heat of a system consisting of a finite
number of bosons confined in an external potential are calculated in different
spatial dimensions.  Using the grand
partition function as the generating function of the partition function, an
iterative scheme is established for the calculation of the partition function
of system with an arbitrary number of particles.  The scheme is applied to
finite number of bosons confined in isotropic and anisotropic parabolic traps
and in rigid boxes.  The specific heat as a function of temperature is studied
in detail for different number of particles, different degrees of
anisotropy, and different spatial dimensions.  The cusp in the specific heat is
taken as an indication of Bose-Einstein condensation (BEC).  It
is found that the results corresponding to a large number of particles are
approached quite rapidly as the number of bosons in the system increases.
For large number of
particles, results obtained within our iterative scheme are consistent with
those of the semiclassical theory of BEC in an external potential based on the 
grand canonical treatment. 

\end{abstract}

\hspace*{0.5 cm}
\section{ Introduction}
Following the foundation laid down by Einstein\cite{1}, discussions on the
phenomena of Bose-Einstein Condensation (BEC) have been of constant
interest to physicists and can be found in standard textbooks on
statistical physics.\cite{2} The conventional treatment is usually based on
the grand canonical ensemble within the context of an ideal Bose gas and
the thermodynamic limit.  The recent beautiful experiments\cite{3} on BEC
in atomic gases, however, lead to the need of a deeper understanding of the
problem of BEC in confined systmes with finite number of bosons.  The fact
that the number of bosons is finite and relatively small in the trap
created experimentally and that the ultracold atomic gas does not
exchange particles with reserviors implies that the canonical ensemble is
more appropriate.\cite{4,5}

It is interesting to study how the thermodynamical quantities such as the
specific heat evolves as the number of particles in the system increases.
In Fermion systems, the differences among results obtained within the
different ensembles have been studied.\cite{f6}  Recently, there have been
some work on finite boson systems in a parabolic confinement within the
context of micro-canonical and canonical ensembles.\cite{4,5}  In the
present work, using the grand partition function as the generating function
of the partition function, an iterative scheme is established for the
calculation of the partition function of system with an arbitrary number of
particles.  The scheme is applied to bosons confined in isotropic and
anisotropic traps and in rigid boxes in arbitrary spatial dimensions.  In
particular, we aim out studying the effects of dimensionality, anisotropy,
and finite number of particles.  Although there exist many
interesting approaches to the problem in the literature,\cite{many}
we concentrate on comparing our results, wherever possible,
with those in the conventional theory and in the semiclassical theory of
Bagnato {\em et al.},\cite{K} which is basically an extension of the
conventional theory to the case with confining potentials.

The plan of the paper is as follows.  In Sec. II, the general formalism and
the recursive scheme are given.  Results for bosons confined in isotropic
and anisotropic parabolic potentials are given in Sec. III.  The
corresponding results for confinement in rigid boxes are given in Sec. IV
together with a discussion on the validity of the integral approximation
commonly used.  A brief summary is given in Sec. V.

\section{Formalism}

We consider a system of $N$ non-interacting bosons.  The partition function
$Z_{N}(v,T)$ is related to the $N$-th derivative with respect to the
fugacity $z$ of the grand partition function $Q_(z,v,T)$ through
\begin{equation}
Z_{N}(v,T)=\frac{1}{N!}
                \frac{\partial^{N}Q(z,v,T)}{\partial z^{N}}
                \mid_{z=0},
\end{equation}
where $T$ is the temperature and $z = \exp(\mu/kT)$ is the fugacity.  Here
$v$
is the variable representing the volume of the system for confinement in a
rigid box and the confining strength in the case of a parabolic confining
potential.  In general, the grand partition function for an ideal
bose gas can be written as\cite{2}
\begin{equation}
Q(z,v,T)=\prod_{p}\frac{1}{1-z e^{-\beta\varepsilon_{p}}},
\end{equation}
where the product is over all the single particle states labelled by $p$
with energies $\epsilon_{p}$, and $\beta = 1/kT$ with $k$ being the
Boltzmann constant.  The zeros of the
chemical potential and the single particle energy are set to be the same.
Combining Eqs.(1) and (2) gives a relation
\begin{equation}
Z_{N}=\frac{1}{N}\sum^{N}_{j=1}
        B_{j}Z_{N-j}
\end{equation}
between the partition function $Z_{N}$ of a $N$-particle system and the
partition functions $Z_{N'}$, with $N' = 1,\cdots, N-1$,
for systems with
less than $N$ particles.
The factor $B_{j}$ is defined as
$B_{j}\equiv\sum_{p} e^{-j\beta\varepsilon_{p}}$.  For simplicity, the
arguments  $v$ and $T$ are left out in the partition functions in Eq.(3).
The internal energy $E$ and the heat capacity $C_{v}$ can then be
calculated using the standard formulae
$E=-\partial \ln Z_{N}/\partial\beta$ and
$C_{v}=\partial E/\partial T$.  These expressions involve the first and
second derivatives of the partition function with respect to temperature.

The first and second derivatives of the partition function of a
$N$-particle system can also be related to quantities in systems with less
particles.  Let $b$ be a dimensionless parameter representing the
temperature.  From Eq.(3), we have
\begin{equation}
\frac{\partial Z_{N}}{\partial b} = \frac{1}{N}\sum_{j=1}^{N}
                \left [
                B_{j} \frac{\partial Z_{N-j}}{\partial b}+
                \frac{\partial B_{j}}{\partial b}Z_{N-j}
                \right ] ,
\end{equation}
and
\begin{equation}
\frac{\partial^{2}Z_{N}}{\partial b^2} = \frac{1}{N}\sum_{j=1}^{N}
                \left [
                B_{j}\frac{\partial^{2}Z_{N-j}}{\partial b^2}+
                2\frac{\partial B_{j}}{\partial b}
                \frac{\partial Z_{N-j}}{\partial b}+
           \frac{\partial^2 B_{j}}{\partial b^2}Z_{N-j}
                \right ] .
\end{equation}
Equations (3)-(5) imply that the partition function and other
thermodynamical quantities
of a $N$-particle system can be calculated from the partition
functions and their derivatives of systems with less particles iteratively
by using the conditions $Z_{0} = 1$, $\partial Z_{0}/\partial b = 0$, and
$\partial^{2}Z_{0}/\partial b^{2} = 0$.  These conditions, when substituted
into Eqs.(3)-(5), give the correct single particle partition function
$Z_{1}$ and its derivatives.  It should be pointed out that these
conditions are independent of the form of the confining potential and spatial
dimension.  The
effects of the confinement are reflected in the energy spectrum and the
explicit form of Eqs.(3)-(5) when they are applied to specific problems.
In what follows, we apply Eqs.(3)-(5) to systems with finite number
of non-interacting bosons confined either in a parabolic potential or in a
rigid box.  In particular, the position of the cusp in the heat capacity as
a function of temperature is taken to be an indication of Bose-Einstein
condensation in the system.  We study the dependence of the position of the
cusp on the number of particles in the system.

\section{ Parabolic Confinement }
\subsection{ Isotropic confinement in 1, 2, and 3 dimensions}

For $M$-dimensional isotropic parabolic confinements characterized by the
energy $\hbar\omega$, the energies of the single particle states are given
by
\begin{equation}
\varepsilon_{n_{i}}=
                \sum_{i=1}^{M}n_{i}\hbar\omega,
\end{equation}
where $n_{i}$ ($i = 1,\cdots,M$) takes on non-negative integers
and
the zero-point energy has been absorbed in the definition of the zero of
energy.  Identifying the dimensionless parameter $b$ as $b \equiv
\exp(-1/\tau)$, where $\tau = kT/\hbar\omega$ is the effective temperature,
and using Eqs.(3)-(5), we have, in $M$ dimensions, the relations
\begin{eqnarray}
Z_{N} & = &\frac{1}{N}\sum_{j=1}^{N}
                \frac{Z_{N-j}}{(1-b^j)^M}, \\
\frac{\partial Z_{N}}{\partial b}&=&\frac{1}{N}\sum_{j=1}^{N}
                \frac{1}{(1-b^j)^M}
                \left [
                \frac{\partial Z_{N-j}}{\partial b}+
                \frac{Mjb^{j-1}}{1-b^j}Z_{N-j}
                \right ] , \\
\frac{\partial^{2}Z_{N}}{\partial b^2}&=&\frac{1}{N}\sum_{j=1}^{N}
                \frac{1}{(1-b^j)^M}
                \left [
                \frac{\partial^{2}Z_{N-j}}{\partial b^2}+
                \frac{2Mjb^{j-1}}{1-b^j}
                \frac{\partial Z_{N-j}}{\partial b} \right.  \nonumber \\
                & & \left .+Mjb^{j-2}\frac{j-1+(jM+1)b^j}{(1-b^j)^2}Z_{N-j}
                \right ].
\end{eqnarray}
Equations (7)-(9), together with the conditions on $Z_{0}$ and its
derivatives, can be used iteratively to obtain $Z_{1}$, $\cdots$,
$Z_{N}$ and their derivatives.  These equations can be easily implemented
numerically for any spatial dimensions.  The heat capacity can be
evaluated using
\begin{equation}
C_{v} = \frac{k}{\tau^{2}} \left[ \frac{b}{Z_{N}} \frac{\partial
Z_{N}}{\partial b} \left ( 1 - \frac{b}{Z_{N}} \frac{\partial
Z_{N}}{\partial b} \right ) + \frac{b}{Z_{N}} \frac{\partial^{2}
Z_{N}}{\partial b^{2}} \right ].
\end{equation}

It should be pointed out that, although exact expression for $Z_{N}$ may
not be possible for general dimensions, exact expressions exist for
$Z_{N}$ and $C_{v}$ in one dimension (1D) and they read
\begin{equation}
Z_{N}^{(1D)}=\frac{1}{\prod_{\ell=1}^{N} (1-b^\ell)},
\end{equation}
and
\begin{equation}
C_{v}^{(1D)}=k\left (
                \frac{\hbar\omega}{kT}
                \right )^2
                \sum_{\ell=1}^{N}
                \frac{\ell^2 b^\ell}{(1-b^\ell)^2}.
\end{equation}
Slightly different expression for $Z_{N}^{(1D)}$ has been obtained by
Brosens {\em et al.}.\cite{4}   The difference
comes from our choice of setting
the zero of energy at the ground state energy of a $m$-dimensional
harmonic oscillator.  This choice, of course, does not affect the result of
the heat capacity.   Figure 1 shows the results of the specific heat
$C_{v}/Nk$ calculated using Eq.(10) as a function of $kT/(\hbar\omega
N^{1/M})$ for systems in $M=1$, $2$, and $3$ dimensions (1D, 2D, 3D) with
different number of bosons.
In Fig. 1(a), it is obvious that the results obtained
numerically using Eqs.(7)-(10) agree well with the results obtained
from the exact expression given by Eq.(12).  This,
in turn, indicates that our
iterative approach is reliable.   In 1D, the specific heat
approaches its large $N$ limit very rapidly in that the curves
corresponding to systems with $N=10$ and $N=10^{4}$ particles are very
close to each other.  As temperature increases, $C_{v}$ approaches its
limiting value of $Nk$.  In contrast to 2D and 3D systems, $C_{v}$ varies
smoothly with temperature and there is no cusp.  This implies that there is
no BEC in 1D system with parabolic
confinement.
This result is consistent with that of the semiclassical treatment of
Bagnato and Kleepner.\cite{K}  These authors studied BEC in traps within
the grand
canonical ensemble and the thermodynamic limit so that the notion of a
density of states is valid.  They found that 1D ideal
Bose gas will display BEC only if the external potential is more confining
than the parabolic form.

Figure 1(b) shows that cusp appears
in the specific heat when the number of particles in the system is large
enough in 2D isotropic parabolic confinement.   The position of the cusp
shifts to higher temperature as the
number of particles increases.   As there will be, strictly speaking, no
sharp transition in a system with finite number of particles, the
appearance of a cusp in the specific heat is taken to be an indication of
BEC.  For system with $N= 10^{4}$ particles, the cusp is at a temperature
$kT/(\hbar\omega N^{1/2}) \approx 0.75$; while the semiclassical
treatment\cite{K}
gives a transition temperature $T_{c}$ at $kT_{c}/(\hbar\omega N^{1/2}) =
\sqrt{6}/\pi \approx 0.7797$ for $N \rightarrow \infty$.  Thus, our results
are consistent with the semiclassical results and we expect the position of
the cusp in $C_{v}$ to approach a value slightly larger than
$kT/(\hbar\omega N^{1/2}) \approx 0.75$ as $N$ increases.

In 3D isotropic parabolic confinement, the specific heat shows a cusp as a
function of temperature for systems with large enough $N$ with the position
of the cusp shifts up in temperature as $N$ increases and approaches a
limit
(see Fig.1(c)).  However, the behaviour is quite different from that in 2D.
As $N$ increases, the cusp becomes sharper and $C_{v}$ decreases rapidly
across the cusp and eventually leads to a discontinuity in $C_{v}$.  For
systems with $N=10^{4}$ particles, the position of the cusp is at
$kT/(\hbar\omega N^{1/3}) \approx 0.9$ and the discontinuity in $C_{v}$ is
about $-6.5Nk$.   According to the semiclassical treatment for BEC in 3D
harmonic traps,\cite{K} the transition temperature is at
$kT_{c}/(\hbar\omega N^{1/3})
\approx 0.912$ for $N \rightarrow \infty$, and there will be a
discontinuity
in $C_{v}$ of the amount given by $[C_{v}(T_{c}^{+}) - C_{v}(T_{c}^{-})]/Nk
\approx -6.57$.  Therefore, our results for large $N$ are consistent with
that of semiclassical theory.  In contrast, the specific heat in $C_{v}$ is
continuous across the cusp in 2D.

\subsection{ Anisotropic confinement in 2 dimensions}

We choose to illustrate the effects of anisotropy by considering
two-dimensional systems.  The energies of the single particle
states in a 2D anisotropic harmonic confinement are
\begin{equation}
\varepsilon_{n_{1},n_{2}}=\hbar\omega_{1}
                (n_{1}+\alpha n_{2}),\ \ \ n_{1}, n_{2}=0,  1, 2, \cdots,
\end{equation}
where $\alpha = \omega_{2}/\omega_{1}$ is the ratio of the frequencies of
the harmonic potentials in the two directions.  In this case, the factor
$B_{j}$ is defined to be
$B_{j}(b)=1/(1-b^j)(1-b^{\alpha j})$, where $b = \exp(1/\tau)$ with $\tau =
kT/\hbar\omega_{1}$.  Equations (3)-(5) can then be applied iteratively to
obtain the heat capacity.  Figure 2 shows the dependence of
the specific heat
on temperature by plotting $C_{v}/Nk$ as a function of $\tau/(\alpha
N)^{1/2} = kT/\hbar(\omega_{1}\omega_{2})^{1/2}N^{1/2}$ for different
values of $\alpha$ with different number of particles in the system.
Figure 2(a) shows the case with fixed degree of anisotropy $\alpha = 80$.
For $N=10^{4}$ particles in such an anisotropic trap, the cusp appears at
$\tau/(\alpha N)^{1/2} \approx 0.71$; while semiclassical theory gives a
value of $0.7797$ for $N \rightarrow \infty$.  For small number of
particles, say $N=10$, $C_{v}$ appears to approach the 1D limit of $Nk$ at
low temperature.  However, as temperature increases, $C_{v}$ increases
again and eventually approaches the 2D limit of $2Nk$ at high temperature.
This behaviour can be understood in that the anisotropy leads to different
energy spectra in the two directions.  At low temperature, only the levels
corresponding to the less confining direction are populated and thus the
system appears to behave in a one-dimensional way.  At higher temperature,
the levels corresponding to the larger confining direction start to become
populated and the system takes on two-dimensional character.  Figure 2(b)
shows the effects of anisotropy for a fixed number of particles $N =
10^{4}$ in the system.  For small $\alpha$, there is no cusp in $C_{v}$ --
a
character of a 1D system.  However, it should be pointed out that even for
small values of $\alpha$, $C_{v}$ approaches the 2D limit of $2Nk$ at high
temperatures.

\section{ Rigid Box}

The iterative approach can also be used to calculate the thermodynamical
quantities of a system of $N$ non-interacting Bosons confined in
a $M$-dimensional rigid box of volume
$v=L^M$. The energies of the single-particle states for a particle of mass
$m$ in such a box are given by
\begin{equation}
\varepsilon_{n_{i}}=\sum_{i=1}^{M}\frac{n_{i}^{2}h^2}{8mL^2},
\end{equation}
where the quantum numbers $n_{i}\ (i=1, 2, \cdots, M)$ takes on positive
integers.  The factor $B_{j}$ in $M$ dimensions is given by
$B_{j}(b)=[\Theta_{j}(b)]^M$ with $\Theta_{j}(b)$ being the factor $B_{j}$
in 1D defined by
$\Theta_{j}(b)\equiv
        \sum_{n=1}^{\infty}
        b^{jn^2} $.
The dimensionless parameter $b$ is taken to be
$b\equiv \exp(-1/\tau)$,  where $\tau$ is the effective temperature given
by  $\tau\equiv 8mkTL^2/h^2$.
In $m$ dimensions, Eqs.(3)-(5) become
\begin{eqnarray}
Z_{N}&=&\frac{1}{N}\sum_{j=1}^{N}
                \Theta_{j}^{M}Z_{N-j}, \\
\frac{\partial Z_{N}}{\partial b}&=&\frac{1}{N}\sum_{j=1}^{N}
                \Theta_{j}^{M}
                \left [
                \frac{\partial Z_{N-j}}{\partial b}+
                \frac{M}{\Theta_{j}}
                \frac{\partial \Theta_{j}}{\partial b}
                Z_{N-j}
                \right ] , \\
\frac{\partial^{2}Z_{N}}{\partial b^2}&=&\frac{1}{N}\sum_{j=1}^{N}
                \Theta_{j}^{M}
                \left \{
                \frac{\partial^{2}Z_{N-j}}{\partial b^2}+
                \frac{2M}{\Theta_{j}^{M}}
                \frac{\partial \Theta_{j}}{\partial b}
                \frac{\partial Z_{N-j}}{\partial b} \right.  \nonumber \\
                & & \left .+ M
                \left [
                \frac{M-1}{\Theta_{j}^2}
                   \left (
                \frac{\partial \Theta_{j}}{\partial b}
                   \right )^2+
                \frac{1}{\Theta_{j}}
                \frac{\partial^2\Theta_{j}}{\partial b^2}
                \right ]
                Z_{N-j}
                \right \}.
\end{eqnarray}
The function $\Theta_{j}(b)$, which is related to the Jacobi's
Theta-function,\cite{math} and its derivatives can be evaluated
numerically by summing the corresponding infinite series to convergence.
The heat capacity can then be obtained using Eq.(10).  Figure 3 shows the
specific heat for different number of bosons in a rigid box in 1D, 2D, and
3D.  Figure 3(a) shows that in 1D rigid box the results for systems with a
small number of particles approach that of a larger number of particles
quite rapidly.  This feature is similar to that in a 1D parabolic trap.
Another feature is that the high temperature limit of $C_{v} = 0.5 k$ is
approached quite slowly.  For example, $C_{v}/Nk$ increases from
approximately $0.478$ to $0.489$ as $\tau/N^{2}$ increases from $20$ to
$80$.

Figure 3(b) shows that in 2D the large $N$ behaviour is approached very
rapidly.  The high temperature limit is approached more rapidly than in the
1D case.  There is no cusp in the specific heat for both 1D and 2D.  This
is consistent with the well-known result that BEC for ideal bose gas
in the absence of external confining potential does not exist in
dimensions lower than three.  Figure 3(c) gives the results in 3D.
The position of the cusp
shifts to approach the large $N$ limit from above as the number of
particles increases.  For $N = 10^{4}$, our numerical results give the
maximum specific heat at the cusp the value $C_{v}/Nk = 2.087$
at the corresponding temperature
$8mkTL^{2}/h^{2}N^{2/3} \approx  0.7556$.
It should be contrasted with the
standard
ideal Bose gas results in 3D in the thermodynamic limit that the maximum
specific heat is $1.925 Nk$ at the transition temperature given by
$8mkT_{c}L^{2}/h^{2}N^{2/3} \approx 0.6713$.
In contrast to the results
corresponding to a 3D parabolic potential, the specific heat is continuous
across the cusp.\par

The infinite series in $\Theta_{j}(b)$ is sometimes approximated
by an integral\cite{2} which yields
$\Theta_{j}\approx (\pi\tau/4j)^{1/2}$.
With this approximation, a suitable dimensionless parameter is
$b\equiv (\pi\tau/4)^{M/2}$.  An
approximate expression of the the factor
$B_{j}(b)$ in $M$ dimensions is
\begin{equation}
B_{j}(b) \approx j^{-M/2}b.
\end{equation}
>From the identity
\begin{equation}
\sum_{n=1}^{\infty}e^{-jn^2/\tau}=
\frac{1}{2}(\frac{\pi\tau}{j})^{1/2}+
(\frac{\pi\tau}{j})^{1/2}
\sum_{q=1}^{\infty}e^{-\pi^2 q^2\tau/j}-\frac{1}{2},
\end{equation}
the integral approximation amounts to retaining only the first term.
Within this approximation, Eqs.(15)-(17) read
\begin{eqnarray}
Z_{N}&=&\frac{1}{N}\sum_{j=1}^{N}
                j^{-M/2}bZ_{N-j}, \\
\frac{\partial Z_{N}}{\partial b}&=&\frac{1}{N}\sum_{j=1}^{N}
                j^{-M/2}
                \left [
                b\frac{\partial Z_{N-j}}{\partial b}+
                Z_{N-j}
                \right ] , \\
\frac{\partial^{2}Z_{N}}{\partial b^2}&=&\frac{1}{N}\sum_{j=1}^{N}
                j^{-M/2}
                \left [
                b\frac{\partial^{2}Z_{N-j}}{\partial b^2}+
                2\frac{\partial Z_{N-j}}{\partial b}
                \right ].
\end{eqnarray}
Within this approximation, the formula for heat capacity is
\begin{eqnarray}
C_{v}(b)&=& \frac{Mk}{2}
        \left [
             \left (
             \frac{M}{2}+1
             \right )
        \frac{b}{Z_{N}}
        \frac{\partial Z_{N}}{\partial b}
           -\frac{M}{2}
           \left (
           \frac{b}{Z_{N}}
           \frac{\partial Z_{N}}{\partial b}
           \right )^2 \right.\nonumber \\
        &  &\left. +\frac{M}{2}
        \frac{b^2}{Z_{N}}
         \frac{\partial^2 Z_{N}}{\partial b^2}
        \right ],
\end{eqnarray}
and the calculated results are shown in Fig.4.
The general features are quite similar to that of the exact results shown
in Fig.3.  It can be shown that the integral approximation gives reliable
results except for systems with only a few particles, in those cases the
approximation may lead to erroneous results.

\section{ Conclusions}

We have studied the BEC of finite number of confined bosons.
An iterative scheme has been used to obtain the
specific heat for systems with different number of
particles confined in an external potential.
As emphasized by van Hove,\cite{van}
no singularities would appear in the partition function and the
thermodynamical quantities of a finite system.
Therefore the cusp of speific heat curve was taken as an indication of
BEC.  Numerical results show that the specific heat
approaches its large $N$ limit rapidly as the number of particles in the
system increases.  Results in different spatial dimensions are obtained.
Results for systems with large $N$, typically $N \approx 10^{4}$ in our
calculations, are consistent with those in the semiclassical theory of
Bagnato {\em et al.}.\cite{K}  As the early experiments have typically
$10^{4}$-$10^{5}$ particles in the trap,\cite{3} we expect that treatments
within the canonical and grand canonical ensembles would give at least
qualitatively consistent results.  Although the iterative scheme is applied
here to Bose systems, it can equally be applied to systems with finite
number of {\em Fermions} confined in an external potential.  In this case,
the variable $B_{j}$ should be modified to be
$B_{j}\equiv \sum_{p}(-1)^{j-1}e^{-j\beta\varepsilon_{p}}$.

\acknowledgments

One of us (PMH) acknowledges the support of a grant from the British
Council under
the UK-HK Joint Research Scheme.  WD is supported by a Postdoctoral
Fellowship of the Chinese University of Hong Kong.



\newpage
\begin{figure}[btp]
\begin{center}
\leavevmode
\epsfbox{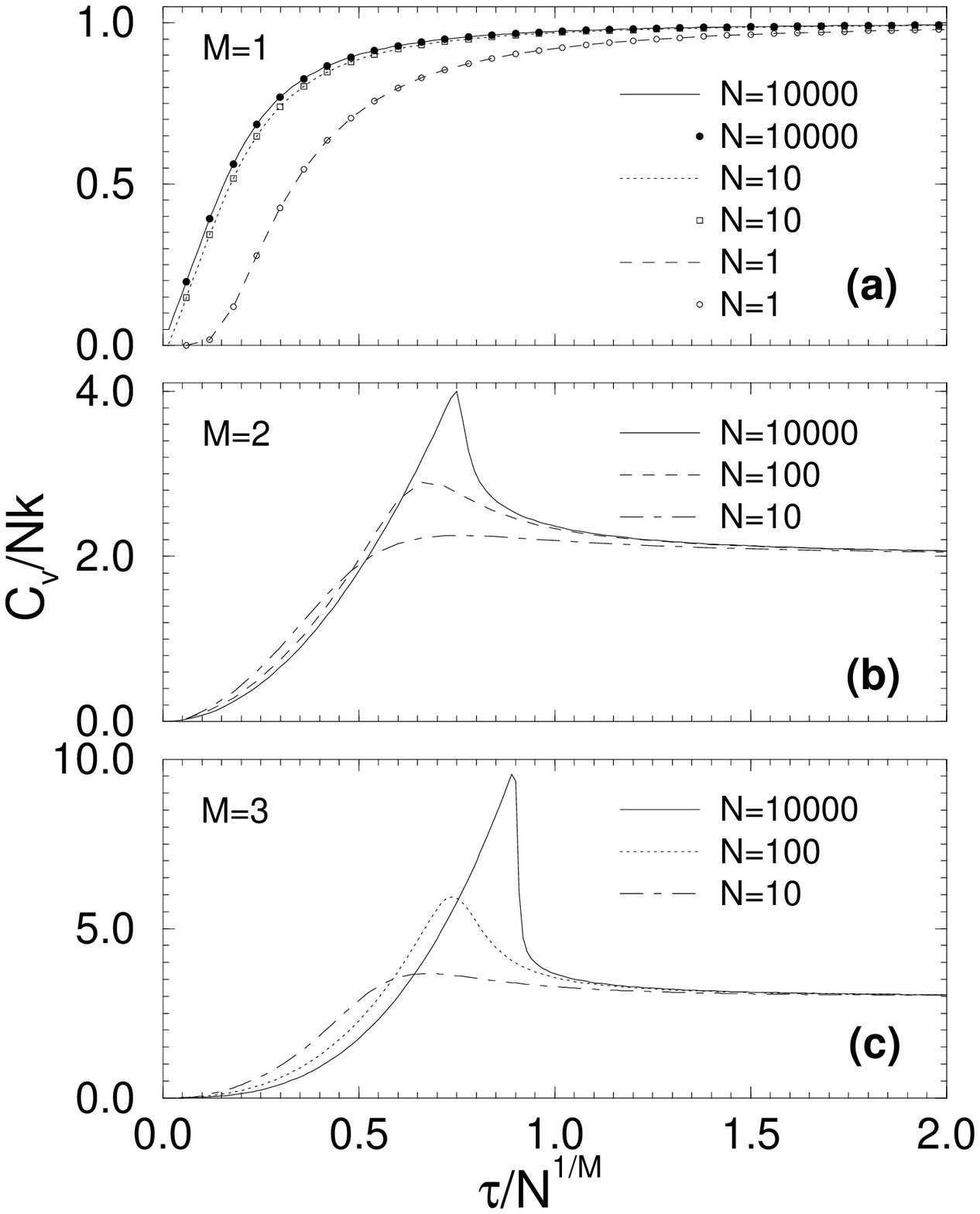}
\end{center}
\end{figure}
\begin{figure}
\caption{The specific heat of a finite number $N$ of bosons confined in an
isotropic parabolic potential as a function of the dimensionless temperature
$\tau/N^{1/M}=kT/(\hbar\omega N^{1/M})$, where $M$ is the spatial dimension for
(a) $M=1$, (b) $M=2$, and (c) $M=3$.  The lines are results of the iterative
scheme for different values of $N$.  The symbols in (a) are the results
obtained using the exact expression Eq.(12).
}\label{Fig. 1}
\end{figure}

\newpage
\begin{figure}[btp]
\begin{center}
\leavevmode
\epsfbox{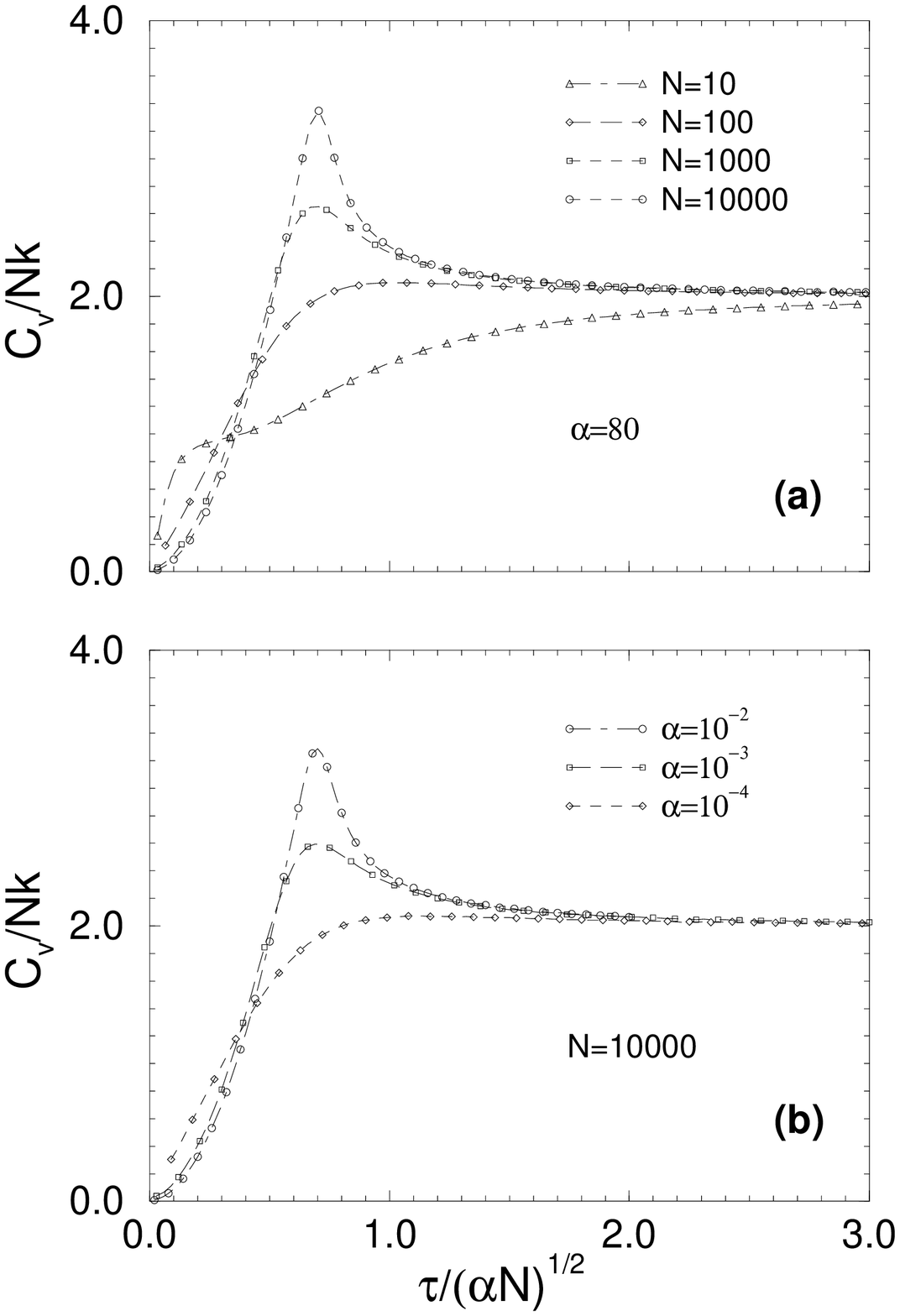}
\end{center}
\end{figure}
\begin{figure}
\caption{ The specific heat of $N$  bosons confined in a
two-dimensional anisotropic parabolic potential as a function of the
dimensionless temperature 
$\tau/(\alpha N)^{1/2}=kT/\hbar(\omega_1\omega_2)^{1/2}N^{1/2}$.
(a) The degree of anisotropy $\alpha$ is taken to be $80$ and the
number of particles takes on four different values of
$N=10$, $10^2$, $10^3$, and $10^4$ .
(b) The number of particle is fixed at $N=10^4$, and the degree of anisotropy
$\alpha$ takes on  $\alpha=10^{-4}$, $10^{-3}$, and $10^{-2}$.
}\label{Fig. 2}
\end{figure}

\newpage
\begin{figure}[btp]
\begin{center}
\leavevmode
\epsfbox{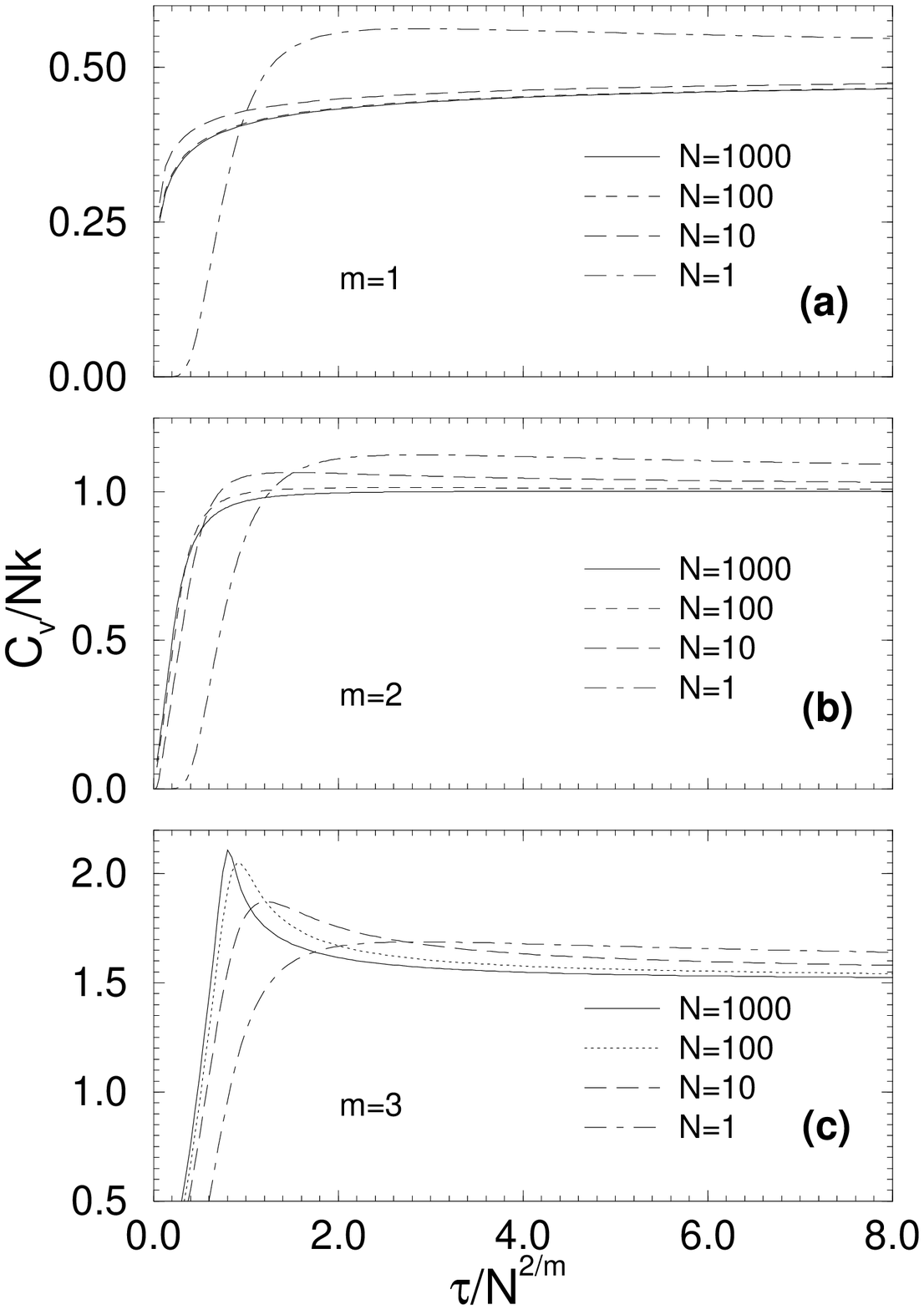}
\end{center}
\end{figure}
\begin{figure}
\caption{The specific heat of $N$ bosons confined in $M$-dimensional rigid box
as a function of the dimensionless temperature
$\tau/N^{2/M}=8mkTL^2/h^{2}N^{2/M}$ for
(a) $M=1$, (b) $M=2$, and (c) $M=3$. Results are obtained using Eqs.(15)-(17)
by summing up the infinite series for $\Theta_{j}(b)$ to convergence.
}\label{Fig. 3}
\end{figure}

\newpage

\begin{figure}[btp]
\begin{center}
\leavevmode
\epsfbox{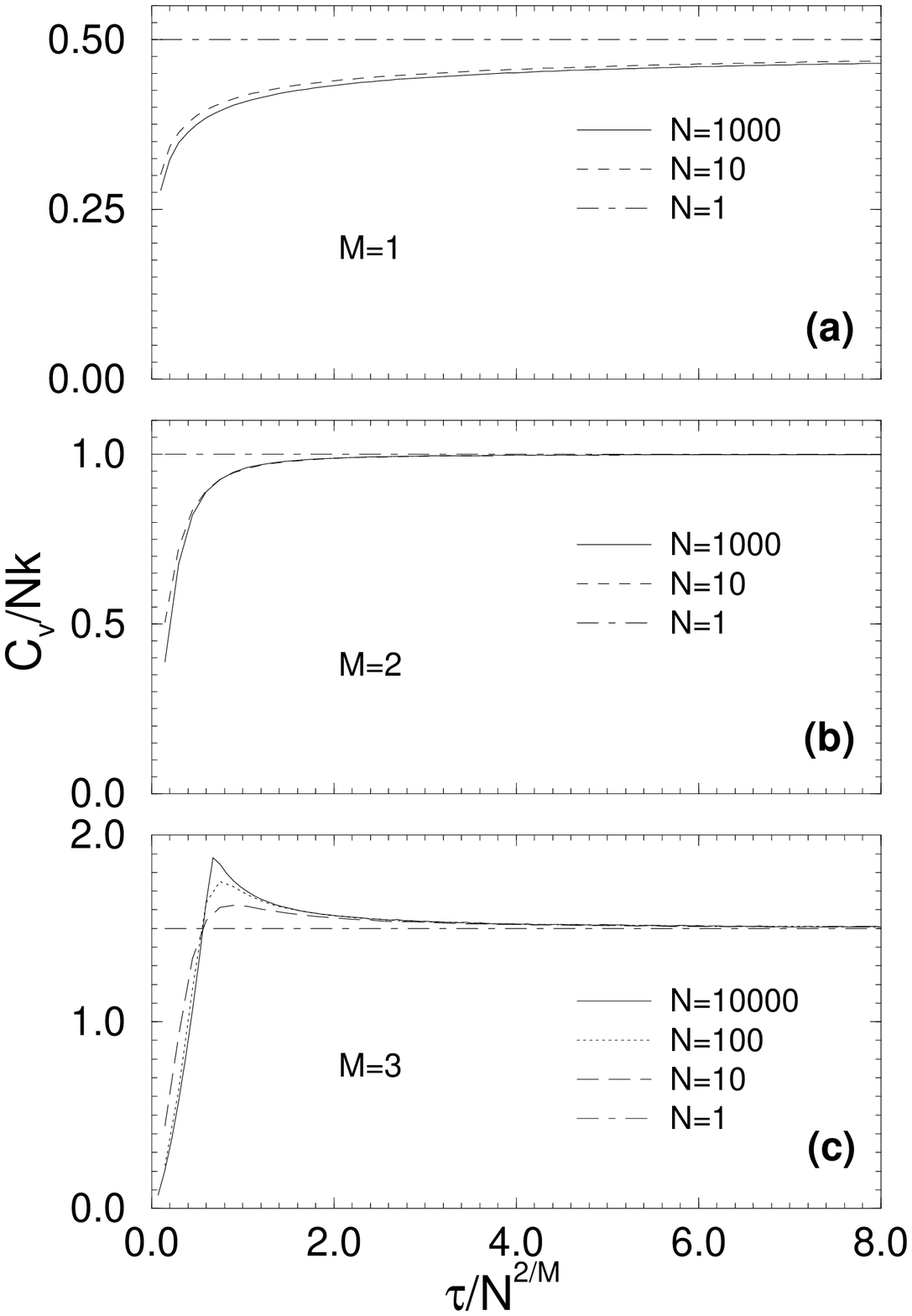}
\end{center}
\end{figure}
\begin{figure}
\caption{The specific heat of $N$ bosons confined in $M$-dimensional rigid box
as a function of the dimensionless temperature $\tau/N^{2/M} =
8mkTL^{2}/h^{2}N^{2/M}$ for (a) $M=1$, (b) $M=2$, and (c) $M=3$.  Results are
obtained using Eqs.(20)-(22) within the integral approximation.
}\label{Fig.4}
\end{figure}


\begin{references}
\bibitem{1} A. Einstein, Ber. Berl. Akad. 261 (1924); {\em ibid.} 3 (1925).

\bibitem{2} See, for example, K. Huang, {\em Statistical Mechanics},
        (Wiley, New York, 1963).

\bibitem{3} M.H. Anderson, J.R. Ensher, M.R. Matthews, C.E. Wieman,
        and E.A. Cornell, Science {\bf 269}, 198 (1995);
        C.C. Bradley, C.A. Sackett, J.J. Tollett, and R.G. Hulet,
        Phys. Rev. Lett. {\bf 75}, 1687 (1995);
        K.B. Davis, M.-O. Mewes, M.R. Andrews, N.J. van Druten, D.S.
Durfee,  D.M. Kurn, and W. Ketterle,
        Phys. Rev. Lett. {\bf 75}, 3969 (1995).

\bibitem{4} F. Brosens, J.T. Devreese, and L. F. Lemmens,
    Solid State Commun. {\bf 100}, 123 (1996).

\bibitem{5} S. Grossmann, and M. Holthaus, Phys. Rev. E {\bf 54},
3495 (1996).

\bibitem{f6} See, for example, R. Denton, B. Muhlschlegel, and D. J. Scalapino,
        Phys. Rev. B {\bf 7}, 3589 (1973);
        H. P. Chen and H. F. Cheung,
        J. Phys.: Condens. Matter {\bf 7}, 6707 (1995) and
        refrences therein.


\bibitem{many}
    S. R. de Groot, G.J. Hooyman, and C.A. ten Seldam
    Proc. R. Soc. A {\bf 203}, 266 (1950);
   D. L. Mills, Phys. Rev. {\bf 134}, A 306 (1964);
    P. K. Pathria, Can. J. Phys. {\bf 61}, 228 (1983);
    K. Kirsten, and D.J. Toms,
    Phys. Rev. A {\bf 54}, 4188 (1996);
    Phys. Lett. A {\bf 222}, 148 (1996);
    Phys. Lett. B {\bf 368}, 119 (1995);
    W. Ketterle and N. J. van Druten,
    Phys. Rev. A {\bf 54}, 656 (1996);
    S. Grossmann and M. Holthaus,
    Phys. Lett. A {\bf 208}, 188 (1995); Z. Phys. B {\bf 97}, 319 (1995);
    Z. Naturforsch. a{\bf 50}, 921 (1995).

\bibitem{K} V. Bagnato, D. E. Pritchard, and D. Kleppner,
        Phys. Rev. A {\bf 35}, 4354 (1987);
    V. Bagnato and D. Kleppner, Phys. Rev. A {\bf 44}, 7439 (1991).

\bibitem{math}See, for example, E. T. Whittaker and G. N. Watson,
        {\em A course of modern analysis}, 4th edition,
        Chapters 20-22, (Cambridge Univ. Press, Cambridge,
        England, 1952).

\bibitem{van} L. van Hove, Physica, {\bf 15}, 951 (1949).


\end{references}
\end{document}